# A wave nature-based interpretation of the nonclassical feature of photon bunching on a beam splitter

Byoung S. Ham
Center for Photon Information Processing, School of Electrical Engineering and Computer Science, Gwangju Institute of Science and Technology 123 Chumdangwagi-ro, Buk-gu, Gwangju 61005, South Korea
bham@gist.ac.kr (submitted on Oct. 12, 2021)

**Abstract:**
Born's rule is key to understanding quantum mechanics based on the probability amplitude for the measurement process of a physical quantity. Based on a typical particle nature of a photon, the quantum feature of photon bunching on a beam splitter between two output photons can be explained by Born's rule even without clear definition of the relative phase between two input photons. Unlike conventional understanding on this matter, known as the Hong-Ou-Mandel effect, here, we present a new interpretation based on the wave nature of a photon, where the quantum feature of photon bunching is explained through phase basis superposition of the beam splitter. A Mach-Zehnder interferometer is additionally presented to support the correctness of the presented method. As a result, our limited understanding of the quantum feature is deepened via phase basis superposition regarding the destructive quantum interference. Thus, the so-called 'mysterious' quantum feature is now clarified by both the definite phase relationship between paired photons and a new term of the phase basis superposition of an optical system.

**Introduction**

One of the most unique features of quantum mechanics is photon bunching on a beam splitter (BS), known as the Hong-Ou-Mandel (HOM) effect [1], where zero coincidence measurement between two output photons results from simultaneously impinging two indistinguishable input photons on a BS [1-6]. Using the particle nature of photons, destructive quantum interference between the reflected and transmitted photon probabilities on a BS is the physical reason of the photon bunching. However, the destructive quantum interference on a BS has not been clearly defined due to the particle nature of photons, and thus the physical origin of photon bunching has been somewhat vague in terms of space-time indistinguishability (independence) between two photons. Although specifying a definite phase of a single photon is prohibited by quantum mechanics, defining a relative phase between two photons is allowed [7]. Here, a completely different approach based on the wave nature of photons is presented to understand the origin of photon bunching on a BS for a general two-input-two-output photon-BS system [8,9]. For this, an analysis of a typical classical example of a one-input-two-output photon-BS system precedes the general two-input-two-output BS system. Moreover, a Mach-Zehnder interferometer (MZI) composed of two BSs is added to validate the present approach.

The wave-particle duality is the bedrock of quantum mechanics, where both natures are mutually exclusive [7,10]. Recently, quantitative measurements of the wave-particle duality between correlated photons have been experimentally observed, where any generated photons are practically mixed states between these two pure properties with a certain weight factor [11]. However, an analytic approach to photon properties in a given system must be treated as either a particle or a wave, where the wave property represents coherence optics for a general classical realm. This is the main reason for avoiding the wave nature when investigating quantum phenomena, especially since the Henbury Brown and Twiss (HBT) experiments for two-photon intensity correlation [12]. HBT opens a new realm of quantum mechanics from the view point of indistinguishable photon characteristics in a space-time domain, where the wave property of a photon is mostly removed since then [13,14]. Based on the indistinguishable characteristics via the particle nature of photons, probability amplitudes via superposition between two photons plays a major role in conventional nondeterministic quantum physics [15-20]. This nondeterministic property of indistinguishable photon characteristics results in 'mysterious' quantum phenomenon such as entanglement between two or more identical (or even different) photons or atoms, because mutual superposition of probability amplitudes has never been clearly defined [21].

To understand this so-called 'mysterious' quantum feature, the well-known HOM effect has been theoretically studied recently regarding the physical origin of photon bunching on a BS [8,9,22]. Using the wave nature of photons, the HOM effect is clearly understood with a relative phase relationship between two input photons. The disappearance of first-order intensity correlation in the HOM effect can be explained by averaging of all different frequency pairs of photons with a wide bandwidth, resulting in a coherence washout effect [22].

As a result, however, the interference fringe of the quantum feature can be retrieved [22] if the bandwidth of entangled photons becomes narrower as observed in refs. 23-26. Even though the concept of phase quantization has been introduced recently in a coupled interferometric system for higher-order quantum correlation corresponding to a N00N state [23], indistinguishability or randomness of probability amplitudes between interacting photons in an optical system via quantum superposition has not been clearly understood [27].

In the present paper, the quantum feature of photon bunching on a BS is analyzed with superposition of quantized phase bases of the BS to investigate the 'mysterious' quantum phenomenon. As a result, four different phase-basis combinations with an equal amplitude probability are taken into account for the same quantum feature. The randomness or indistinguishability of finding photons is now clearly understood via the newly defined superposed phase bases of a given optical system. Moreover, the quantum phenomenon of photon bunching on a BS via coincidence measurements is now understood through quantized phase bases of the optical system [9], resulting in indistinguishable photon characteristics in a space-time domain in terms of the photon probabilities. Thus, the present wave nature-based analysis opens the door to coherence quantum optics based on phase quantization of an optical system. Successful understanding of the 'mysterious' quantum feature gives us a definite clue as to why a photon should take only two paths even in a multi-slit interferometer [28-31]. In other words, Born's rule of amplitude probability of a photon is now easily understood as a result of not only photon characteristics but also specific optical systems interacting with the photons.

**Results**

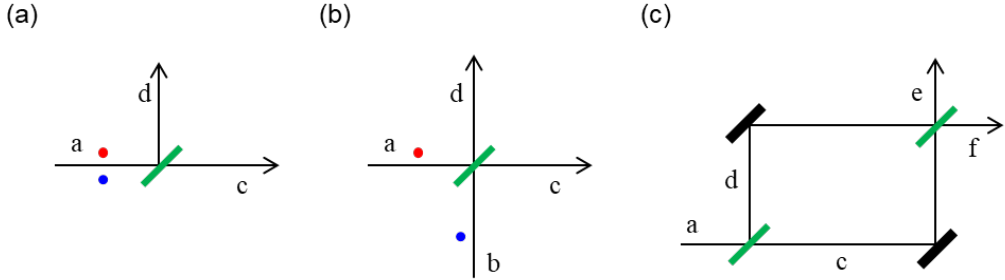

Fig. 1. Schematics of BS interacting with single photons. (a) For the same input port. (b) For the different input ports. (c) Mach-Zehnder interferometer with the same input. The red and blue dots indicate input photons.

Figure 1 shows two-photon impinging BS models, where Fig. 1(a) is for a single input port and Fig. 1(b) is for double input ports. Figure 1(c) is an extension of Figs. 1(a) or 1(b) for an MZI having two consecutive BSs. The input photons are denoted by red and blue dots to separate them for quantum feature analysis. Specifying a relative phase between two input photons does not violate quantum mechanics as discussed in the EPR paradox [21] and Bell inequality violation [32], where two spin bases in EPR are replaced by two phase bases of a BS, $\varphi_{BS} \in \left\{-\frac{\pi}{2}, \frac{\pi}{2}\right\}$ [8,9,27]. Assigning a different phase basis to different input photons randomly results in indistinguishable characteristics for the photon-optics system. Here, the wave nature of a photon is thoroughly considered to investigate the quantum feature of photon bunching on a BS.

According to coherence optics, the BS in Fig. 1 is represented by a matrix form [33]:

$$[BS]_\pm = \frac{1}{\sqrt{2}}\begin{bmatrix} 1 & \pm i \\ \pm i & 1 \end{bmatrix}, \qquad (1)$$

where the sign '$\pm$' directly comes from a lossless and balanced (50/50) nonpolarizing BS, and '$i$' represents a $\frac{\pi}{2}$ phase shift between the reflected and transmitted photons in the form $E_d = E_c e^{\pm i\pi/2}$. For the present analysis based on the wave nature of a photon, coherent input photons are individually applied for either the same or opposite phase bases of $\varphi_{BS}$. In Fig. 1(a), the phase difference between two input photons, however, does not affect the results of the output photons. In other words, the predetermined $\pm\frac{\pi}{2}$ phase relationship between the output photons of a BS for the one-input photons is an inherent feature of the BS regardless of their phases.

(i) Analysis for Fig. 1(a)

A-1. For the same $[BS]_\pm$ applied to each input photon from the same input port

$$\begin{bmatrix} E_c \\ E_d \end{bmatrix} = \frac{E_0}{\sqrt{2}} \begin{bmatrix} 1 & \pm i \\ \pm i & 1 \end{bmatrix} \begin{bmatrix} \sqrt{2} \\ 0 \end{bmatrix}$$
$$= E_0 \begin{bmatrix} 1 \\ \pm i \end{bmatrix}. \tag{2}$$

Thus, $E_c = E_0$ and $E_d = \pm i E_0$, where the corresponding output intensities are $I_c = I_d = I_0$ ($|E_0|^2$). Here, $I_0$ is the intensity of each input photon. Equation (2) represents the balanced outputs in both output ports 'c' and 'd' by the 50/50 nonpolarizing BS. To make sense of this, the BS physics reflects not the intensity but the amplitude of a photon's path selections. This amplitude probability of a photon interacting with a BS has already been clarified by Born's rule [28-31], where the related Sorkin's parameter (limiting the path choice up to two) [34] is not due to the photon's characteristics but instead the BS's inherent phase bases [33]. The phase quantization of the photon-optics system has been recently discussed regarding emerging physics in quantum mechanics [27]. Equation (2) also represents typical understanding of a single photon as a particle nature interacting with a BS: $\langle I_c \rangle = \langle I_d \rangle = I_0$.

A-2. For the opposite $[BS]_+$ and $[BS]_-$ combination applied to each input photon separately

$$\begin{bmatrix} E_c \\ E_d \end{bmatrix} = \frac{E_0}{\sqrt{2}} \left\{ \begin{bmatrix} 1 & i \\ i & 1 \end{bmatrix} \begin{bmatrix} 1 \\ 0 \end{bmatrix} \pm \begin{bmatrix} 1 & -i \\ -i & 1 \end{bmatrix} \begin{bmatrix} 1 \\ 0 \end{bmatrix} \right\}$$
$$= \sqrt{2} E_0 \begin{bmatrix} 1 \\ 0 \end{bmatrix} \text{ or } i\sqrt{2} E_0 \begin{bmatrix} 0 \\ 1 \end{bmatrix}. \tag{3}$$

Thus, the output photons have two choices either $E_c = \sqrt{2} E_0$ & $E_d = 0$ for the symmetric case of basis superposition or $E_c = 0$ & $E_d = i\sqrt{2} E_0$ for the antisymmetric case. Thus, the corresponding output intensities are either $I_c = 2I_0$ & $I_d = 0$ or $I_c = 0$ & $I_d = 2I_0$, respectively, at an equal probability. From equation (3), the mean value of photon measurements in each output port is uniform, i.e., $\langle I_c \rangle = \langle I_d \rangle = I_0$. However, the mean value of coincidence measurements is always zero: $\langle R_{cd} \rangle = 0$. This relationship represents the non-classical feature of photon bunching on a BS regarding the one-input port case due to destructive quantum interference between the opposite phase bases of the BS [8]. Although equation (3) is an unexpected and unprecedented result in both coherence and quantum optics with unentangled photons, the nonclassical feature in Fig. 1(a) still satisfies the classical lower bound of $g^{(2)}(t=0) = 0.5$ due to the equal probability of basis combinations in equations (2) and (3) [8,35].

A typical misunderstanding of photon bunching in the quantum optics community relates to intensity probability being based on each photon's random path selections, resulting in contraction to the MZI directionality due to the bunched photon by the first BS, resulting in equal splitting in the second BS. A more severe misunderstanding of the photon bunching phenomenon is with regard to the independency between two input photons, where independent photon-based HOM effects using independent light sources are actually for phase-locked photons via an optical method such as an etalon [6], an optical cavity [25], same pulse pumping [3-5], or heralded beating-signal gating [36].

(ii) Analysis for Fig. 1(b)

B-1. For the same $[BS]_\pm$ applied to each input photon from different input port,

$$\begin{bmatrix} E_c \\ E_d \end{bmatrix} = \frac{E_0}{\sqrt{2}} \begin{bmatrix} 1 & \pm i \\ \pm i & 1 \end{bmatrix} \begin{bmatrix} 1 \\ e^{i\theta} \end{bmatrix}$$
$$= \frac{E_0}{\sqrt{2}} \begin{bmatrix} 1 \pm i e^{i\theta} \\ \pm i + e^{i\theta} \end{bmatrix}, \tag{4}$$

where $\theta$ is the difference phase between two input photons. Thus, $E_c = \frac{E_0}{\sqrt{2}} (1 \pm i e^{i\theta})$ and $E_d = \frac{E_0}{\sqrt{2}} (\pm i + e^{i\theta})$, where the corresponding intensities are $I_c = I_0 (1 \mp \sin\theta)$ and $I_d = I_0 (1 \pm \sin\theta)$. Here, the phase $\theta$ is a controllable parameter as in Young's double-slit experiments, resulting in the uniform intensity over $\theta$ in each output port, $\langle I_c \rangle = \langle I_d \rangle = I_0$. If $\theta = \pi/2$, both output intensities are either $I_c = 0$ & $I_d = 2I_0$ or $I_c = 2I_0$ & $I_d = 0$. If $\theta = -\pi/2$, the results are $I_c = 2I_0$ & $I_d = 0$ or $I_c = 0$ & $I_d = 2I_0$. In both cases, the mean photon intensity in each output port becomes uniform $\langle I_c \rangle = \langle I_d \rangle = I_0$. However, the mean coincidence measurements between two output photons are always zero regardless of the θ choices. These two specific cases of θs represent the nonclassical conditions of photon bunching, satisfying the HOM effect [8]. The uniform mean output intensity is not due to random phases between the input photons [22], but instead the phase basis superposition of the BS according the present wave nature of photons.

B-2. For opposite BS matrices applied to each photon from different input port

$$\begin{bmatrix} E_c \\ E_d \end{bmatrix} = \frac{E_0}{\sqrt{2}} \left\{ \begin{bmatrix} 1 & i \\ i & 1 \end{bmatrix} \begin{bmatrix} 1 \\ e^{i\theta} \end{bmatrix} \pm \begin{bmatrix} 1 & -i \\ -i & 1 \end{bmatrix} \begin{bmatrix} 1 \\ e^{i\theta} \end{bmatrix} \right\} \frac{1}{\sqrt{2}}$$
$$= E_0 \begin{bmatrix} 1 \\ e^{i\theta} \end{bmatrix} \text{ or } iE_0 \begin{bmatrix} e^{i\theta} \\ 1 \end{bmatrix}. \quad (5)$$

Thus, the results are either $E_c = E_0$ & $E_d = e^{i\theta}E_0$ for the symmetric case or $E_c = ie^{i\theta}E_0$ & $E_d = iE_0$ for the antisymmetric case, where the corresponding output intensities are $I_c = I_d = I_0$, regardless of $\theta$. The mean coincidence measurement is $\langle R_{cd} \rangle = 1$, violating the nonclassical feature of photon bunching. As analyzed for Fig. 1(c) below, however, this opposite BS phase basis combination does or does not satisfy experimental results depending on the input port choices for either Fig. 1(b) or 1(a) for an MZI, respectively [37,38]. Satisfaction with equation (5) represents independent input photons, which is not possible for the one-input port-based MZI. In other words, equation (5) is applicable only to phase independent paired photons, and thus photon bunching of the HOM effect must be limited to phase correlated photon pairs as analyzed in ref. 8. More fundamental understanding on this violation matter originates in the indistinguishability or randomness of finding photons in both paths of MZI, where satisfaction with equation (5) contradicts this superposition principle of self-interference resulting from Born's rule (see Discussion) [38].

(iii)  Analysis for Fig. 1(c) with the same input photons

In this case of MZI with one input port and two coherent photons, i.e., $E_a = \sqrt{2}E_0$ and $E_b = 0$, the matrix representation for Fig. 1(c) is presented for different combinations of the BS phase bases.

C-1. For the same BS matrix to each photon with equation (2)

$$\begin{bmatrix} E_e \\ E_f \end{bmatrix} = \frac{1}{\sqrt{2}} \begin{bmatrix} 1 & \pm i \\ i & \pm 1 \end{bmatrix} \begin{bmatrix} E_c \\ E_d \end{bmatrix}$$
$$= \frac{E_0}{\sqrt{2}} \begin{bmatrix} 1 & \pm i \\ \pm i & 1 \end{bmatrix} \begin{bmatrix} 1 \\ \pm i \end{bmatrix}. \quad (6)$$

Thus, $E_e = 0$ and $E_f = \pm i\sqrt{2}E_0$, where the corresponding intensities are $I_e = 0$ and $I_f = 2I_0$. As a result, equation (6) satisfies the general solution of photon directionality of MZI optics, where no path length difference is assumed [37,38]. This result corresponds to equation (4) with $\theta = \pm\frac{\pi}{2}$ resulting from the first BS in Fig. 1(c).

C-2. For opposite BS matrix to each photon with equation (5)

$$\begin{bmatrix} E_e \\ E_f \end{bmatrix} = \frac{1}{2} \left\{ \begin{bmatrix} 1 & i \\ i & 1 \end{bmatrix} \begin{bmatrix} E_c \\ E_d \end{bmatrix} \pm \begin{bmatrix} 1 & -i \\ -i & 1 \end{bmatrix} \begin{bmatrix} E_c \\ E_d \end{bmatrix} \right\}$$
$$= E_0 \begin{bmatrix} 1 \\ e^{i\theta} \end{bmatrix} \text{ or } iE_0 \begin{bmatrix} e^{i\theta} \\ 1 \end{bmatrix}, \quad (7)$$

where $\begin{bmatrix} E_c \\ E_d \end{bmatrix} = E_0 \begin{bmatrix} 1 \\ e^{i\theta} \end{bmatrix}$ for the symmetric case, and $\begin{bmatrix} E_c \\ E_d \end{bmatrix} = iE_0 \begin{bmatrix} e^{i\theta} \\ 1 \end{bmatrix}$ for the antisymmetric case. As a result, $E_e = E_0$ & $E_f = e^{i\theta}E_0$ for the symmetric case or $E_e = ie^{i\theta}E_0$ & $E_f = iE_0$ for the antisymmetric case. The corresponding intensities are $I_e = I_f = I_0$ for both cases, regardless of $\theta$. With an equal probability in both cases, the overall mean value of output intensities in each output port is also $\langle I_c \rangle = \langle I_d \rangle = I_0$, regardless of θ. Because equation (7) is satisfied with one input port-based MZI, this conclusion definitely violates the general solution of MZI directionality in both classical and quantum regimes [37,38]. This is the reason why the opposite BS phase-basis combination in equation (5) for Fig. 1(b) should be prohibited. This result corresponds to equation (5).

**Discussions**
Regarding Fig. 1(b) having two input ports, the resulting nonclassical feature of photon bunching was satisfied only for the same basis combinations of the BS matrices if and only if their phase difference is fixed at either $\theta_+ = +\frac{\pi}{2}$ or $\theta_- = -\frac{\pi}{2}$. Different basis combinations must be applied to phase independent case in a two-input MZI, resulting in uniform output intensity in each output port (not shown). Thus, photon bunching on a BS is originated in the destructive quantum interference via the same basis superposition for different input photons in equation (4) or different phase basis combination for the same input photons in equation (3). This understanding reminds us the quantum operator-based destructive interference for photon bunching on a BS. However, the no phase relation between two input photons in conventional quantum mechanics are now clearly demonstrated in the present analysis with a particular difference phase. This same conclusion with different details is not obvious even though the wave-particle duality is the fundamental bedrock of quantum mechanics. Here, the particular difference phase of $\theta_\pm = \pm\frac{\pi}{2}$ is the same as the phase basis of a BS, where entangled photons inherently satisfy this condition as analyzed in refs. 8 and 22. The mean photon measurements in individual output ports of the BS are always equal to each other in all cases, satisfying the general solution of coherence optics whose measurement is based on ensemble averages. This destructive interference between phase-basis combinations is the unique finding in this paper for the physical origin of the photon bunching. Here, it must be clarified that this BS phase basis superposition must be correlated with the relative phase θ between input photons for the quantum feature [39]. This understanding implies a coherence-quantum feature of macroscopic quantum correlation via phase basis randomness [40].

**Conclusion**
In conclusion, the quantum feature of photon bunching on a BS was analytically demonstrated using the wave nature of quantum mechanics for both one and two input ports. For this, analysis was conducted using the coherence optics-based phase basis of a BS. With tensor products of two input ports of a BS for two input photons, a total of four different combinations of a coupled system were analyzed. Unlike the conventional approach based on the particle nature of a photon, the photon bunching phenomenon on a BS for one input port was demonstrated via the amplitude probability of the phase bases of the BS. This nonclassical feature was not explained by coherence (classical) optics dealing with ensemble average but by quantum optics using coincidence measurements.

In the two-input photon-BS system, the nonclassical feature of photon bunching via coincidence measurements was fully satisfied for the same phase basis combination but not for the opposite basis choice, resulting in the fundamental condition of input photons via a particular phase relationship. Like the operator-based quantum mechanical analysis for the HOM effect, destructive interference between phase basis-correlated input photons is the physical origin of the nonclassical feature. However, ensemble averages in each output port of all proposed combinations made the nonclassical feature disappear, satisfying classical physics. Thus, a measurement technique of coincidence detection via destructive interference between photon-BS amplitude probabilities reveals the nonclassical feature, where the essential requirement between two input photons from different input ports is $\pm\frac{\pi}{2}$ phase difference. Thus, the so-called 'mysterious' quantum feature is nothing but a special case of coherence optics for destructive quantum interference via a specified measurement technique in the time domain.

Unlike the common mistakes of intensity probability and photon independency to quantum optics community regarding the BS physics, the nonclassical feature of photon bunching on a BS was clarified for the phase basis superposition of correlated photon pairs. Even though the measurements by coincidence detections are equal to the conventional interpretation based on the particle nature of a photon, the present analysis based on the wave nature opens the door to a deeper and clearer understanding of the fundamental physics of quantum phenomenon, where photon bunching is no longer strange.

**Acknowledgment**
BSH acknowledges that this work was supported by GIST-GRI 2021 and ICT R&D program of MIST/IITP (2021-0-01810). This work was motivated by the discussions with Prof. J. Lee at Hanyang University, Seoul, Korea, regarding Fig. 1(a).